\documentclass[%
 aip,
 jmp,
 amsmath,amssymb,
 reprint,%
]{revtex4-2}

\usepackage{graphicx}
\usepackage{dcolumn}
\usepackage{bm}

\usepackage[utf8]{inputenc}
\usepackage[T1]{fontenc}
\usepackage{mathptmx}
\usepackage{etoolbox}

\usepackage{xcolor} 
\usepackage{soul} 
\usepackage{xfrac} 
\newcommand{\Tc}{$T_{\mathrm{c}}$}
\newcommand{\Tb}{$T_{\mathrm{b}}$}
\newcommand{\Rn}{$R_{\mathrm{n}}$}
\newcommand{\Psat}{$P_{\mathrm{sat}}$}
\renewcommand{\deg}{$^{\circ}$}
\newcommand{\degC}{$^{\circ}$C}
\newcommand{\um}{$\mu$m}

\newcommand{\eg}{\emph{e.g.}}

\makeatletter
\def\@email#1#2{%
 \endgroup
 \patchcmd{\titleblock@produce}
  {\frontmatter@RRAPformat}
  {\frontmatter@RRAPformat{\produce@RRAP{*#1\href{mailto:#2}{#2}}}\frontmatter@RRAPformat}
  {}{}
}%
\makeatother
\begin{document}

\preprint{AIP/123-QED}

\title[Hafnium-based TES bolometers]{Development of hafnium-based transition edge sensor bolometers for cosmic microwave background polarimetry experiments}
\author{K.~M.~Rotermund}
\email{KRotermund@lbl.gov}
\affiliation{Lawrence Berkeley National Lab, United States of America}
\author{X.~Li}
\affiliation{Lawrence Berkeley National Lab, United States of America}
\author{R.~Carney}
\affiliation{Lawrence Berkeley National Lab, United States of America}
\author{D.~Yohannes}
\affiliation{SEEQC, United States of America}
\author{R.~Cantor}
\affiliation{Star Cryoelectronics, United States of America}
\author{J.~Vivalda}
\affiliation{SEEQC, United States of America}
\author{A.~Chambal-Jacobs}
\affiliation{SEEQC, United States of America}
\author{A.~Suzuki}
\affiliation{Lawrence Berkeley National Lab, United States of America}

\date{\today}

\begin{abstract}
Next generation cosmic microwave background (CMB) polarimetry experiments aim to deploy order $\sim$500,000 detectors, requiring repeatable and reliable fabrication process with stable and uniform transition edge sensor (TES) bolometer performance. 
We present a hafnium (Hf)-based TES bolometer for CMB experiments. We employ a novel heated sputter deposition of the Hf films enabling us to finely tune the critical temperature (\Tc) between 140~mK -- 210~mK. We found elevated deposition temperatures result in films with lower stress, larger crystal sizes, and a smaller relative abundance of the m-plane to c-plane $\alpha$ phase, all contributing to the empirical linear dependence of critical temperature on deposition temperature. Crucially, the heated sputter deposition simultaneously ensures that the critical temperature does not drift despite exposure to heat throughout ongoing fab processes as long as the initial deposition temperature is not exceeded. \Tc's lower than 170~mK require deposition temperature greater than 400\degC, far in excess of typical temperatures the wafer may experience. This ample thermal budget allows us to relax the stringent thermal management that conventional aluminum manganese (AlMn) TES bolometers require, for which temperatures as low as 200\degC\ -- 250\degC\ are used to anneal the AlMn in an effort to adjust the \Tc. Hf additionally exhibits an intrinsic steep superconducting transition (we measure $\alpha>200$) and a corresponding high loop gain (exceeding $\mathcal{L}>10$ deep in the transition). We precisely design the normal resistance of the TES to range between 10 milli-Ohm and 1 Ohm through an interdigitated geometry, making these TES bolometers compatible with both time-division, frequency-division, and microwave multiplexing readout systems. We report on bolometer parameters including critical temperature, normal resistance, I-V curves, saturation power, time constant, and loop gain. \\
\end{abstract}

\maketitle

\section{\label{sec:intro}Introduction}


The cosmic microwave background (CMB) has remarkable spatial uniformity with temperature anisotropies on the order of $\Delta T/T\sim10^{-5}$.\cite{Smoot1992} The CMB's polarization is several orders of magnitude weaker, consequently requiring experiments with significantly higher sensitivities.\cite{Abazajian2016} Modern CMB polarimetry experiments make use of transition edge sensor (TES) bolometers, a technology at such an advanced stage that they are photon-noise limited.\cite{Abitbol2017} Increased sensitivities are therefore primarily achieved by increasing the number of detectors. The number of detectors has increased for each generation of CMB polarimetry experiments. The ambitious next step from current (third generation) to future (fourth generation) experiments will see an order of magnitude increase from $16,000-22,000$ detectors (\eg\ SPT-3G, \textsc{Bicep} Array, Simons Array, Simons Observatory) to $500,000$ detectors. \cite{Abitbol2017} Such a dramatic scaling requires detector fabrication process with large throughput, consistency, and high uniformity between runs. 

Consistency and uniformity of the TESs 
is paramount to this endeavour. Aluminum manganese (AlMn) is an alloy that has been studied extensively and is the superconducting metal of choice for TES bolometers, in large part because several factors can be tuned to achieve the desired critical temperature (\Tc) and normal-state resistance (\Rn). \cite{Li2016} While several groups have successfully fabricated and deployed AlMn TESs, the CMB community has experienced challenges in achieving a repeatable and stable \Tc. Slight drifts of the deposition environment that may occur between runs can result in deviations in the film properties. 
To mitigate this, a reference wafer is commonly used for each run to test the AlMn properties. Furthermore, strict thermal management needs to be adhered to throughout the fabrication process as the \Tc\ of AlMn is sensitive to exposure to heat. Despite these challenges, AlMn TESs have been successfully deployed on Advanced ACTPol, \cite{Li2016} \textsc{Polarbear}-2, \cite{Westbrook2018} and Simons Observatory. \cite{Stevens2020}

Here we report our findings on an alternative superconducting metal, hafnium (Hf). In our development of a Hf-based TES, we design for typical CMB detector parameters and ensure compatibility with current detector fabrication efforts such that they may be easily adopted. Several current and next generation CMB polarimetry experiments operate at a bath temperature (\Tb) of around 100~mK. We adopt this set-point and consider further TES bolometer parameters accordingly. 
Additionally, we design for normal-state resistances of order 10~m$\Omega$ and 1~$\Omega$, ensuring their versatility across different multiplexing techniques.

This work is presented in the following way: In Sect. \ref{sec:hf} we introduce Hf and its superconducting properties. Sect. \ref{sec:design} presents key design parameters for a Hf TES followed by details of the fabrication process in Sect. \ref{sec:fab}. Testing results and analysis are given in Sect. \ref{sec:results}. We conclude in Sect. \ref{sec:conclusion}.

\section{Superconductivity of Hafnium}\label{sec:hf}

Hafnium oxide is a commonly used dielectric in complementary metal-oxide-semiconductor (CMOS) applications as its dielectric constant is four to six times higher than that of the more conventionally known silicon oxide.\cite{Wilk2001} While hafnium oxide is a high-$\kappa$ dielectric, elemental hafnium is a superconductor. Elemental hafnium is an attractive superconducting film for the low-\Tc\ detector community for many reasons, including that its bulk \Tc\ is near 128~mK,\cite{Kraft1998} the London penetration depth is estimated to be 20~nm,\cite{Kraft1998} and the surface kinetic inductance is high at around $15-20$~pH/$\square$ for a 125~nm film,\cite{Coiffard2020} thus making it well-matched to needs across many experiments.
Previous detector efforts include TES calorimeters, \cite{Lita2009,Hunacek2018,Safonova2024} microwave kinetic inductance detectors (MKIDs) sensitive to optical and near-infrared photons, \cite{Zobrist2019,Coiffard2020} superconducting tunnel junctions, \cite{Kraft1998,Kim2012} as well as quantum parity detectors \cite{Ramanathan2024} more recently.

\subsection{Heated sputter deposition}

\begin{figure}
\includegraphics[width=0.7\textwidth]{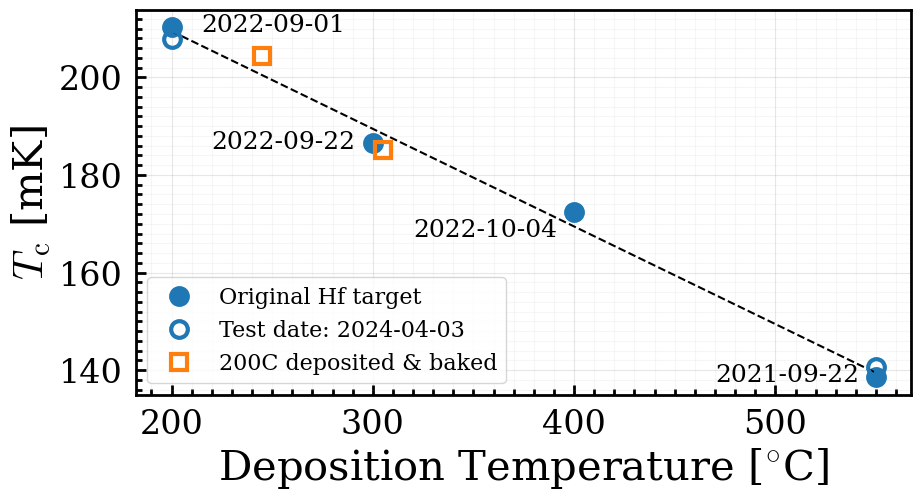}
\caption{\label{fig:heatedsputter} The critical temperature of hafnium is linearly dependent on the deposition temperature. \Tc\ drops as the deposition temperature is increased. The solid data points represent average \Tc's of four or more samples at the given deposition temperature. 
}
\end{figure}

Past efforts on Hf detectors used sputter depositions without heating the substrate. Various approaches were used to target the desired \Tc, including but not limited to, varying the film thickness, adjusting the plasma power during deposition, and using the proximity effect with other metals.\cite{Lita2009,Coiffard2020,Safonova2024} 
In contrast, we heat the wafer during the deposition of our films. The wafer is heated between 200\degC\ and 550\degC. A consistent 247~nm of Hf are deposited for all our wafers. We find a linear dependence of the critical temperature on the deposition temperature, see Fig.~\ref{fig:heatedsputter}, where the \Tc\ drops 
as the deposition temperature is increased. On average, every additional 10\degC\ results in 2~mK drop in \Tc. This allows us to finely tune the \Tc\ by adjusting the deposition temperature. 

The data in Fig.~\ref{fig:heatedsputter} represents critical temperatures as measured on both bare Hf films as well as TES-patterned bolometers. We find no significant deviations in \Tc\ before and after patterning beyond the 5~mK distribution in \Tc\ across and between wafers.  

Hf's \Tc\ further shows stability in time. Solid data points were measured within a few weeks (up to a couple of months) after the initial deposition. The hollow blue circles were re-tested 1.5 and 2.5 years after deposition, stored under ambient conditions, and exhibit no change within our typical 5~mK scatter to \Tc.

\subsubsection{Annealing effects on \Tc}



\begin{figure}
\includegraphics[width=0.65\textwidth]{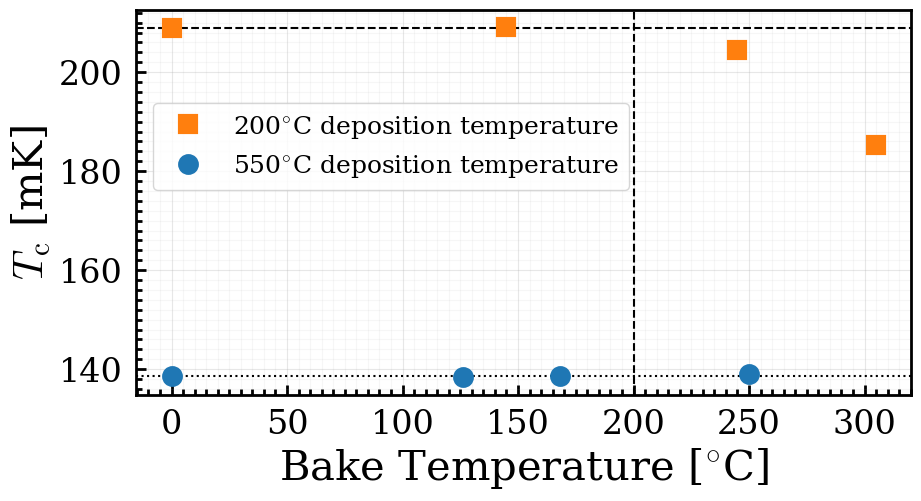}
\caption{\label{fig:Hfbake} Effect on hafnium's critical temperature when baked post deposition. $T_c$ is unchanged as long as the bake temperature remains below the initial deposition temperature. $T_c$ lowers slightly when the bake temperature exceeds the initial deposition temperature.}
\end{figure}

A further property that makes Hf such an attractive TES material is its robustness against exposure to heat. We baked $\sim1$~cm$^2$ chips (Hf originally deposited at 200\degC\ and 550\degC) in atmosphere on a hot plate for 30~min. The bake temperatures ranged from 125\degC\ $-$ 305\degC. Fig.~\ref{fig:Hfbake} shows the measured \Tc\ values as a function of bake temperature. For the chips originally deposited at 550\degC, the \Tc\ remains unchanged for all bake temperatures. In contrast, the \Tc\ is unchanged while the bake temperature remains below 200\degC\ for the chips originally deposited at 200\degC. The \Tc\ then drops as the original deposition temperature is exceeded during the bake. 
The new \Tc\ coincides with the expected \Tc\ for a film that was originally deposited at this elevated temperature, see the hollow orange squares in Fig.~\ref{fig:heatedsputter}. Hf deposited with a heated sputter deposition is therefore robust against further heat exposure as long as the initial deposition temperature is not exceeded. 
For \Tc's below 170~mK, the deposition temperature must be greater than 400\degC. 
Typical temperatures reached during micro-fabrication processes do not exceed $\sim350$\degC\ (\eg\ PECVD) unless the wafer is explicitly heated, we therefore have a substantial thermal budget.

This result is in stark contrast to the behaviour of AlMn. Li et al. (2016) found annealing temperatures of 200\degC\ -- 250\degC\ to have a substantial affect on the \Tc. For different film thickness (ranging from 120~nm to 345~nm), an increase in bake temperature of 50\degC\ resulted on average in an increase in \Tc\ of $\sim$130~mK. In contrast, a similar 50\degC\ bake of a Hf chip (originally deposited at 200\degC) resulted in a drop in \Tc\ of $\sim$19~mK. The \Tc\ of Hf can therefore be tuned with about seven times more precision than AlMn. AlMn is more sensitive to modest heat exposure compared to Hf with an outsized effect. Due to the high variability in the \Tc\ of AlMn from run to run, current AlMn TES development commonly incorporates an intentional bake of the AlMn to adjust and fine-tune the \Tc\ after deposition. Notably, while annealing AlMn causes the \Tc\ to increase, it has the opposite effect on that of Hf.


\subsubsection{Crystal orientation of hafnium}

\begin{figure*}
\includegraphics[width=0.95\textwidth]{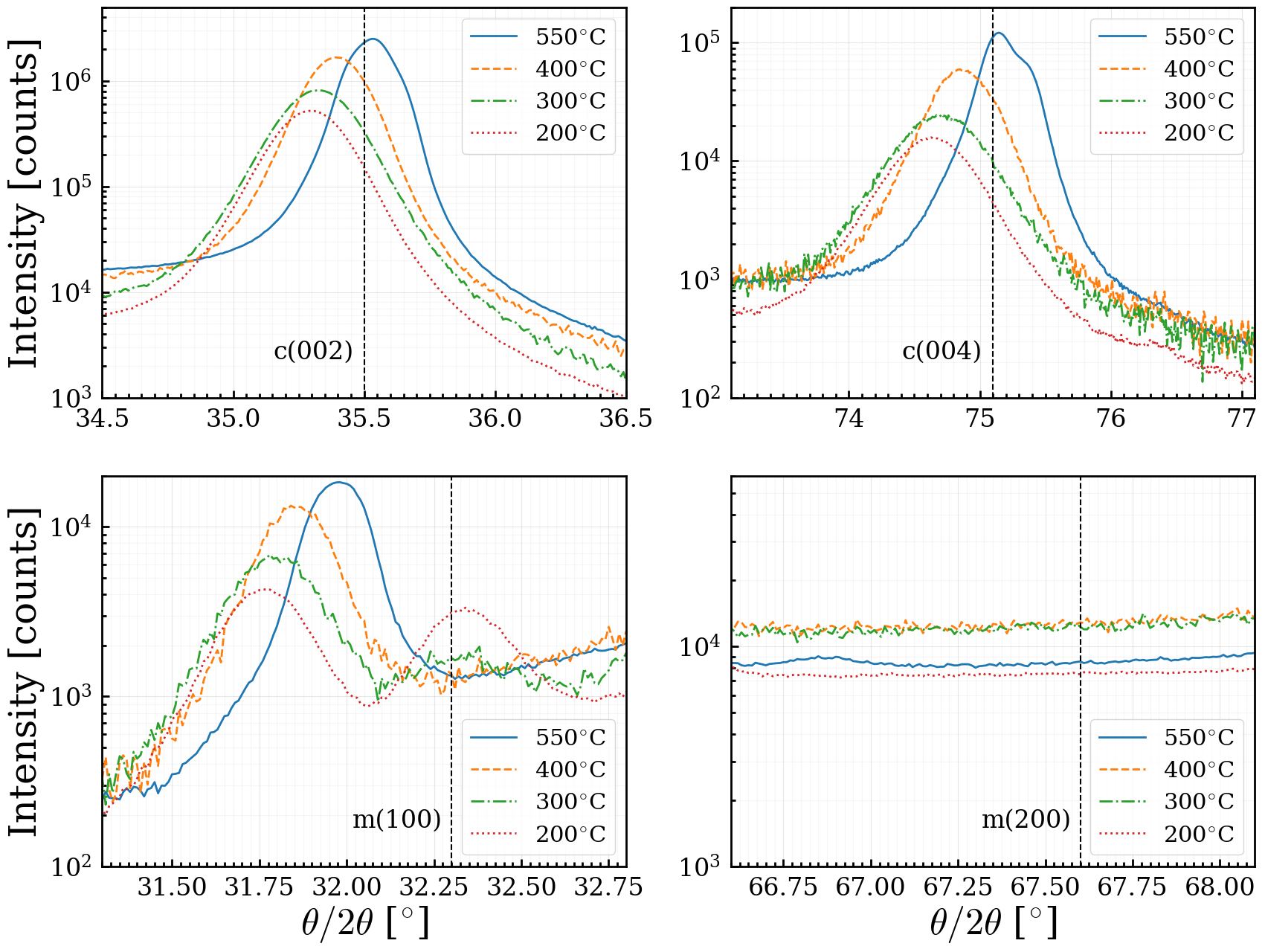}
\caption{\label{fig:xrdLBNL} XRD measurements on Hf film deposited at 200\degC, 300\degC, 400\degC, and 550\degC. The $\alpha$ c-plane (002) and the matching, second-order c-plane (004) Bragg peaks of Hf are present in both films ({\it{top}} row). The $\alpha$ m-plane (100) phase is present in the 200\degC\ and 300\degC\ film ({\it{bottom left}} panel). However, the matching, second-order m-plane (200) is not detected in any films ({\it{bottom right}} panel).}
\end{figure*}

To understand how the critical temperature of Hf changes with deposition temperature we conducted x-ray diffraction (XRD) measurements of samples deposited at 200\degC, 300\degC, 400\degC, and 550\degC. The XRD measurements were done in parallel beam geometry, in which the angle of the incoming beam is fixed at a low 1.5\deg\ angle.
The detector is scanned from 20\deg\ $-$ 80\deg. 
The peaks associated with the $\alpha-$phase of Hf are shown in Fig.~\ref{fig:xrdLBNL}. 

The $\alpha$ c-plane (002) and the matching, second-order c-plane (004) Bragg peaks of Hf are located at 35.467\deg\ and 75.074\deg, respectively. 
The downward shift of the peaks for films deposited at progressively lower temperatures suggests a compressive film, while the 550\degC\ film is nearly stress-free. Additionally, the full width half max (FWHM) of the 200\degC\ film is larger than that of the 550\degC\ film, indicating that the 200\degC\ film has smaller crystal size. 
We also observe the $\alpha$ m-plane (100) at 32.316\deg\ in the 200\degC\ and 300\degC\ films with decreasing intensity. 
The second-order m-plane (200) phase is not detected.




In summary, we find that a higher deposition temperature results in 1) larger crystal sizes, 2) lower-stress films, and 3) lower abundances of the m-plane relative to the c-plane $\alpha$ phase. These three material properties likely contribute to a lower \Tc. Our finding that larger crystal size and lower-stress film contribute to lower \Tc\ is corroborated by work done by Coiffard et al. (2020) on room temperature deposited Hf. They found a decreasing \Tc\ to coincide with larger crystal grains and less compressive film stress. A lower-stress film contributing to lower \Tc\ is additionally corroborated by Hunacek et al. (2018), who found as similar correlation. Lastly, the lower abundances of the m-plane relative to the c-plane $\alpha$ phase affecting the \Tc\ was also observed by Coiffard et al. (2020), who found a higher \Tc\ to coincide with an increase in the (002)-plane to (010)-plane ratio.



\subsubsection{Residual resistance ratio}



\begin{figure}
\includegraphics[width=0.65\textwidth]{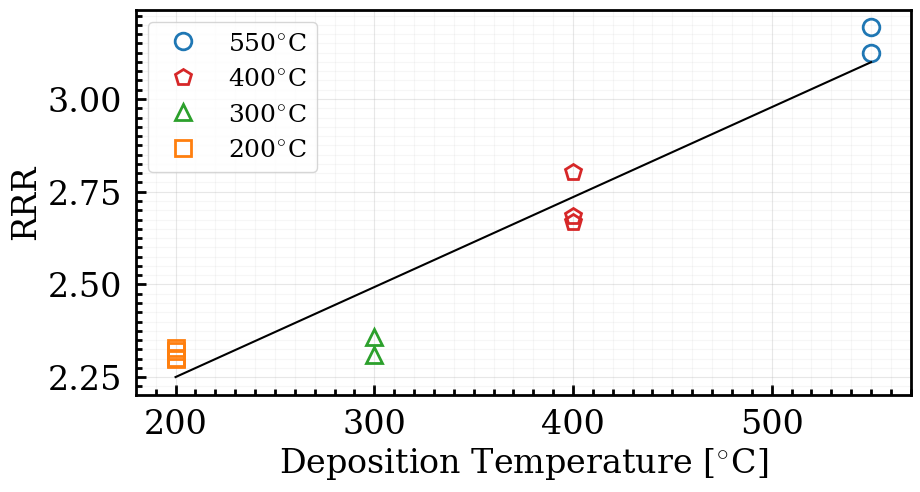}
\caption{\label{fig:RRR} The residual resistance ratio of hafnium increases linearly with higher deposition temperatures.}
\end{figure}

We also measured the residual resistance ratio (RRR) as a function of deposition temperature. We define RRR as the 4~K resistance to its normal resistance (RRR $=R_{4\mathrm{K}}/$\Rn). We find a linearly increasing dependence of the RRR on the deposition temperature, as seen in Fig.~\ref{fig:RRR}. We attribute the higher RRR to both the lower film stress and larger crystal size. 


\section{Design of a hafnium-based TES bolometer}\label{sec:design}


One guiding principle in developing a Hf-based TES is compatibility with ongoing CMB polarimetry detector efforts. Changes to the design and fabrication process should be minimal such that a Hf-based TES can be easily adopted and integrated. The largest difference to current AlMn TES fabrication process flow is the stage at which the TES is deposited and patterned. 
As we are using a heated sputter deposition, the Hf needs to be deposited first on the wafer. AlMn does not have this requirement. 



\subsection{Interdigitated TES}



The dramatic increase of the number of detectors used for each generation of CMB polarimetry experiments has necessitated a corresponding increase in the scalability of the readout of the detectors. Multiplexing several detectors on a single readout line is key to limiting the thermal load on the cold stage of the instrument where the detectors are located, as well as reducing complexity and cost of the system. There are several methods in which the readout can be multiplexed and this is one key area in which CMB polarimetry experiments differ from each other. 

For time-division multiplexing (TDM) readout techniques, the normal resistance of the TES are typically small -- on the order of 10~m$\Omega$. 
CMB-S4 is a next generation polarimetry experiment that has opted to use a TDM readout scheme.
%
In contrast, the normal resistance of the TES can be relatively large for frequency-division multiplexing (FDM) readout techniques -- on the order of 1~$\Omega$. 
The satellite experiment LiteBIRD is a notable future experiment that will use a FDM readout scheme.

Both TDM and FDM are readout schemes that operate at low frequencies (MHz range) and are mature technologies. At high frequencies (GHz range) a newer approach has been developed -- microwave multiplexing ($\mu$-mux). 
$\mu$-mux schemes have the distinct advantage or being capable of multiplexing far greater number of detectors than either TDM or FDM techniques. Similar to TDM, the normal resistance of the TES must be on the order of 10~m$\Omega$. 
\textsc{Bicep}/Keck,\cite{Cukierman2020} AliCPT,\cite{Salatino2021} and Simons Observatory\cite{Jones2024} are all experiments that have either demonstrated or are actively employing a $\mu$-mux readout scheme on sky.



We measured the square resistance of a 247~nm thin film Hf to be $\approx800~\mathrm{m}\Omega$. This places it in the suitable ballpark for a FDM readout scheme. To reduce the normal resistance to $\mathcal{O}(10~\mathrm{m}\Omega)$ for TDM and $\mu$-mux readout systems, the width of the TES must be approximately 100 times larger than the length. In an effort to maintain a compact TES footprint, we use an interdigitated design, see Fig.~\ref{fig:olympus}. Interdigitated TES bolometers have been successfully deployed by QUBIC, \cite{Marnieros2020} a CMB polarimetery experiment, and are being explored by the x-ray and gamma-ray spectroscopy community.\cite{Ullom2015} With the interdigitated approach we can use the TES design to target a precise normal resistance across two orders of magnitude without changing the fabrication process. 

\subsection{Dual-\Tc\ TES}

A TES bolometer optimized for on-sky measurements will saturate immediately in the laboratory as the photon load on a wafer is substantially larger in the laboratory as compared to on-sky measurements once deployed. To enable laboratory testing on a science grade wafer, dual-\Tc\ bolometers are designed, as done previously by \textsc{Polarbear}\cite{Suzuki2012} and \textsc{Bicep}2.\cite{Ade2015} A dual-\Tc\ bolometer has two transitions at two different critical temperatures -- a science and calibration TES. The \Tc\ for optical testing in the lab must be higher to accommodate the larger optical loading. To ensures that the science and calibration transitions do not interfere with each other, we target $T_{\mathrm{c,calib}}>400$~mK. We use titanium (Ti) as the superconducting material for the calibration TES 
and employ a similar interdigitated geometry of the eletrical leads to design for the desired normal resistance. We target a normal resistance for the calibration TES that is about two to four times larger than the science normal resistance.

\section{Fabrication Process}\label{sec:fab}
 
The fabrication process is a collaborative academia--industry effort that builds on the partnerships established by previous detector efforts of.\cite{Suzuki2022} The detector design is created at Lawrence Berkeley National Laboratory. Silicon wafers (150~mm diameter, 500~\um\ thick, 100 crystal orientation) 
are sourced from Rogue Valley Microdevices.\footnote{https://roguevalleymicrodevices.com/} Star Cryoelectronics deposits the hafnium.\footnote{https://starcryo.com/} The remaining micro-fabrication steps are conducted by Seeqc.\footnote{https://seeqc.com/} Lastly, the wafer is diced by GDSI.\footnote{https://www.dieprepservices.com/} 

A TES bolometer measures the power of incoming photons and must be thermally isolated from the bath temperature. This is achieved by suspending the bolometer on a thin membrane with a weak thermal link to the wafer substrate held at the bath temperature. We therefore source Si wafers with both a 2.0~\um\ low-stress SiN and a 0.45~\um\ SiO$_2$ layer on the front- and back-side.

We begin by blanketing the wafer with 247~nm of Hf deposited with a heated sputter deposition. The Hf is then etched to define the TES using a chlorine plasma etch. 
Next, the Hf is sputter-cleaned for 10~min to thoroughly remove the native oxide that grows on Hf. This step is crucial for good electrical contact, as was also found by Lita et al. (2009). Photoresist is deposited and lithographed followed by 300~nm of niobium nitride (NbN) through a sputter-deposition and subsequent lift-off, providing superconducting leads to the TES. 400~nm of low-stress SiN are deposited to passivate the Hf-NbN TES. 

We proceed with a sputter etch to remove oxide growth on exposed NbN at via sites in the SiN passivation layer, followed immediately by a sputter deposition of 450~nm of Nb, creating a Nb ground plane for the detector RF circuitry as well as providing superconducting electrical leads for the Ti-based calibration TES. A chlorine plasma etch 
is used to pattern the Nb. Again, a sputter etch to remove oxides precedes the sputter-deposition and lift-off for the Ti layer defining the calibration TES. Next, we perform a nitrogen (N$_2$) pre-clean prior to depositing 570~nm of silicon-rich SiN using plasma-enhanced chemical vapor deposition (PECVD). This is the SiN microstripline of the detector RF circuitry and also passivates the Ti-based TES. Vias are plasma etched into the SiN using a CHF$_3$ + O$_2$ chemistry. 
A second 650~nm Nb layer is deposited creating the Nb microstripline for the RF circuitry, again using a sputter clean to remove oxides, a sputter deposition, and a chlorine plasma etch. 
A 92~nm film of PdAu is e-beam evaporated and deposited followed by a lift-off for the RF termination resistor for the CMB signal. The last layer to be deposited is 500~nm of palladium (Pd) using e-beam evaporation. The volume of the Pd increases the heat capacity of the bolometer island and is used to regulate the time-constant of the TES bolometer.

\begin{figure*}
\includegraphics[width=0.95\textwidth,trim={0cm 6.8cm 7cm 0cm},clip]{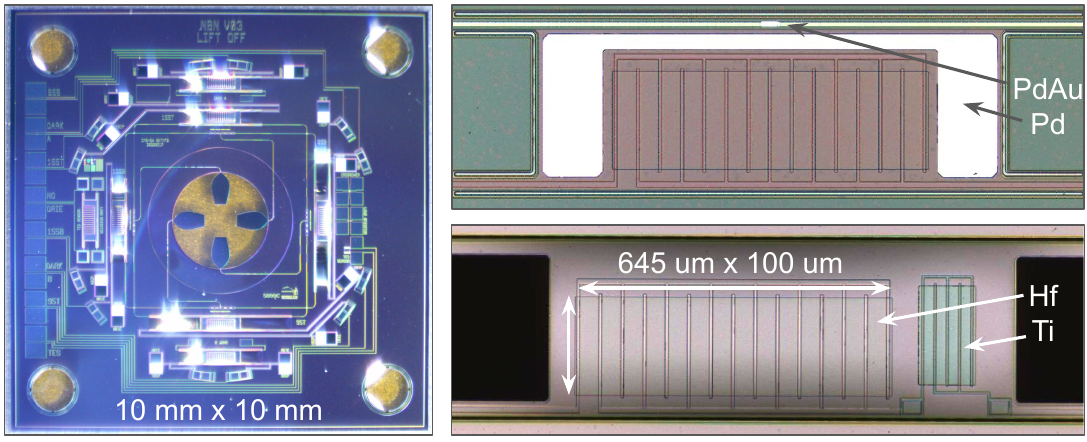}
\caption{\label{fig:olympus} Photographs of the completed full-stack CMB detector wafer with Hf-based TES integrated into the pixel. \emph{Left:} AMScope image of a full 10~mm $\times$ 10~mm detector pixel. \emph{Right:} Olympus confocal microscope on $10\times$ magnification. The Hf is the inner rectangle ($645~\mu\mathrm{m} \times 100~\mu\mathrm{m}$) with NbN interdigiated leads. \emph{Top:} Interdigitated Hf-based TES with a PdAu RF termination resistor and Pd volume for additional heat capacity. \emph{Bottom:} Test structure including an interdigitated Hf-based science TES and an interdigitated Ti-based calibration TES in series on the same released bolometer island.}
\end{figure*}

Next we perform a front-side etch of the SiN + SiO$_2$ films immediately on top of the Si wafer, delineating the bolometer legs. 
The entire front-side of the wafer is coated in 2~\um\ of photoresist to protect the wafer during a backside deep reactive ion etch (DRIE) to release the bolometers and antennas. The wafer is cleaned with an O$_2$ plasma etch. Lastly, the wafer is diced using a stealth dicing technique, in which a laser is focused on the inside of the wafer, breaking the Si bonds and therefore singulating the chips cleanly without disturbing the sensitively released membranes. Microscope images of the fully fabricated detector pixel with a Hf-based science TES and a Ti-based calibration TES are shown in Fig. \ref{fig:olympus}.

\subsection{Superconducting leads}\label{sec:TESleads}

We use superconducting leads to connect to the TES to minimizing parasitic series resistances. Achieving a reliable electrical connectivity to the Hf proved challenging. Several common superconductors were considered -- Nb, Al, and NbN.

\begin{figure*}
$\vcenter{\hbox{\includegraphics[width=0.6\textwidth,trim={0cm 2.8cm 8cm 0cm},clip]{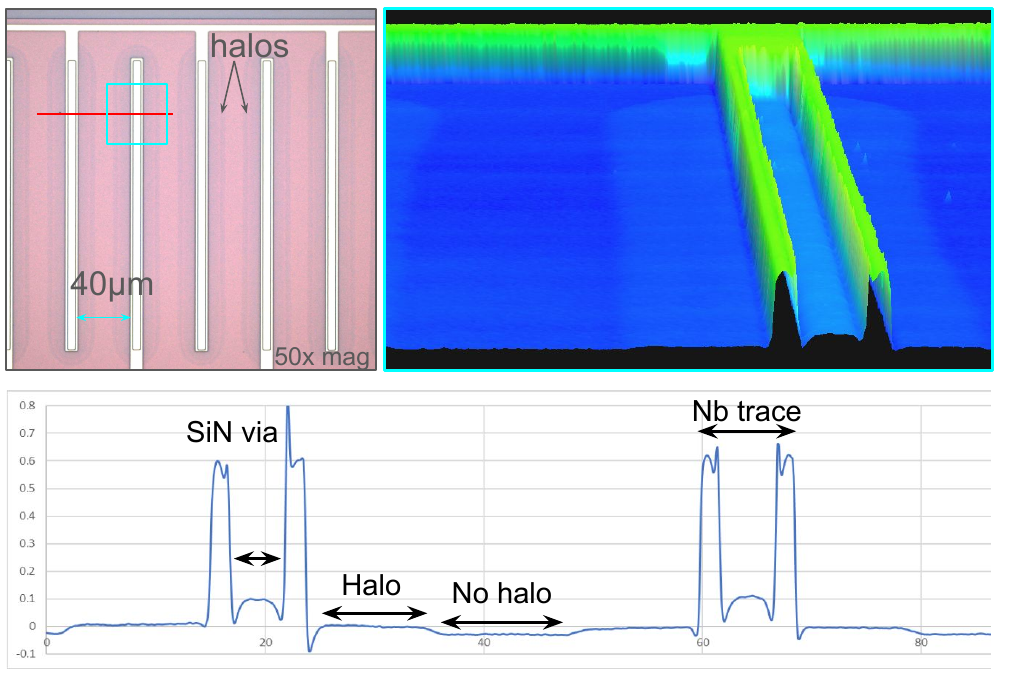}}}$
$\vcenter{\hbox{\includegraphics[width=0.35\textwidth]{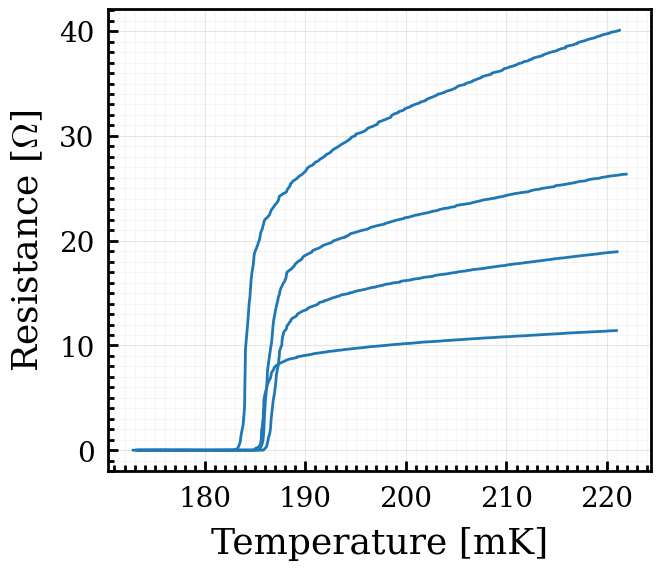}}}$
\caption{\label{fig:halos} \emph{Left:} TES test structure with Nb leads. Olympus confocal microscope on $50\times$ magnification identifying the halos around the SiN vias. \emph{Top-left:} Color image zoomed in on a section of the interdigitated TES design. The pink is the Hf, white traces are the Nb leads, small vertical rectangles with the Nb leads delineate the SiN vias. Darker pink halos are visible around each SiN via. \emph{Top-middle:} 3-dimensional side profile of one Nb trace (the location of which is identified with a cyan rectangle). The color scale represents height. The dark blue region is the Hf, with a clear increase in height (lighter blue) visible around the Nb trace. \emph{Bottom-Left:} Height profile across two Nb leads, identified by the red line in the color image. \emph{Right:} Four-point resistance measurement for test structures with Al leads. Different curves represent the same test structures from across the wafer.}
\end{figure*}

Initially, niobium with a critical temperature of ~9.2~K was selected as the superconducting lead. Both Nb and Hf are etched with a chlorine plasma etch, so the original TES fab process consisted of 1) patterning and etching the Hf, 2) passivating the Hf with SiN and etching vias through the SiN, and finally 3) depositing and etching Nb. Nb is deposited through a sputter deposition during which the wafer is water cooled. Before both the SiN and Nb deposition, the exposed Hf is sputter cleaned to remove any hard native oxide. HfO$_2$ is an insulator and detrimental to the performance of the TES. With careful removal of the HfO$_2$, we made promising electrical contact between the Hf and Nb. The TES critical temperature and normal resistance behaved as expected. However, the yield across the wafer was poor, $<50\%$. 
A majority of the TESs did not transition into their supderconducting state. This coincided with a visible discoloration around the vias in the SiN where the Nb and Hf made contact, see Fig. \ref{fig:halos}. These ``halos'' start as small, irregular bubbles around the vias and grow in size until they merge and form symmetric halos. Using a confocal microscope we measured the height profile and determined the halos to be elevated with respect to the unaffected Hf. 

The cause of the halos is still undetermined. If the halos are the growth of HfO$_2$, then the increased height may be attributed to the larger volume of the oxide relative to the elemental Hf. Similarly, some Hf-Nb alloy may be forming near the interface. However, when the same fabrication process was performed on cleaved wafer quadrants, the halos did not appear. A focused ion beam (FIB) lift-out measurement would provide an elemental analysis across the interdigiated TES with fine enough resolution to determine the properties of the halos. The difference between the full wafer and quarter wafer results, may suggest a film stress. In this scenario, a large film stress may cause the Hf to delaminate from the wafer substrate, resulting in elevated regions. 

To alleviate any potential stress-induced problems, we considered superconducting leads that could be deposited with a lift-off process. Here we used aluminum with a critical temperature of 1.2~K. The switch to the lift-off method eliminated the need to etch the lead material and so the fab process no-longer includes a SiN layer between the two superconductors. However, SiN is deposited after both the Hf and Al to passivate the TES. With the Al leads, no halos were seen on any TES bolometers, substantially increasing the yield across the wafer. Unfortunately, the superconducting transition of the Hf was impacted by the Al. While the initial transition from the superconducting to normal state is still steep, as the TES resistance approaches its normal state, the transition becomes more gradual. In fact, a defined plateau is generally not observed and the resistance continues to slowly increase, as can be seen in the \emph{right} panel of Fig. \ref{fig:halos}. This is likely due to the abnormally long coherence length of Al. 
Aluminum's bulk coherence length is 1600~nm, compared to that of Nb at 38~nm.\cite{Kittel2005} 
While a slow roll-off is not detrimental to the TES performance as long a sharp transition in which the TES can be biased remains, it does pose a challenge to defining the normal resistance. We found the anticipated $1/N$ scaling of the interdigiated design to no longer hold with no discernible relation of the measured normal resistances.

Most recently, we switched to using niobium nitride (with a critical temperature of 16.5~K\cite{Hazra2016}) as the superconducting leads. NbN is also deposited as a lift-off process. The fabrication process flow remains similar to that of Al, with full details outlined in the prior section. As a result of switching to NbN leads, a steep superconducting transition is once-again observed with a sharp plateau to the normal state, similar to un-patterned Hf. The expected $1/N$ scaling from the interdigiated design is also recovered. Fig. \ref{fig:TcRn} shows the results of four-point resistance measurements of the Hf-based TES with NbN leads. Additionally, no halos are observed and the yield across the wafer is high. While testing statistics are low, all test devices measured behaved well. Further test results of Hf-based TES bolometers with NbN superconducting leads are presenting in the following section.


\section{Results}\label{sec:results}


Testing of the completed wafer is done in a Bluefors LD-400 dilution refrigerator (DR) capable of reaching 10~mK temperatures. Mounted to the 4~K stage are six Quantum Design DC SQUIDS with which the TES bolometers can be characterized. A bias circuit is connected to each SQUID and bolometer pair, where the bias resistance is either 0.5~m$\Omega$ or 20~m$\Omega$, depending on the normal resistance of the TES under test. 
The SQUID current is measured in response to one of three variables: 1) A sweep in bath temperature to measure \Tc. 
2) The SQUID current response as a function of the voltage across the TES to measure I-V curves from which the normal resistance, parasitic resistance,  and saturation power can be determined. 
3) Lastly, we apply a square waveform and monitor the decay time in response to the stimulus and determine the associated loop gain. 

In addition to measuring the full suite of TES bolometer properties, test structures on singulated dies can be connected to two LakeShore 372 AC resistance bridges to measure \Tc, \Rn, and $\alpha$ from four-point resistance measurements. 

\subsection{Critical temperature}

\begin{figure}
\includegraphics[width=0.44\textwidth]{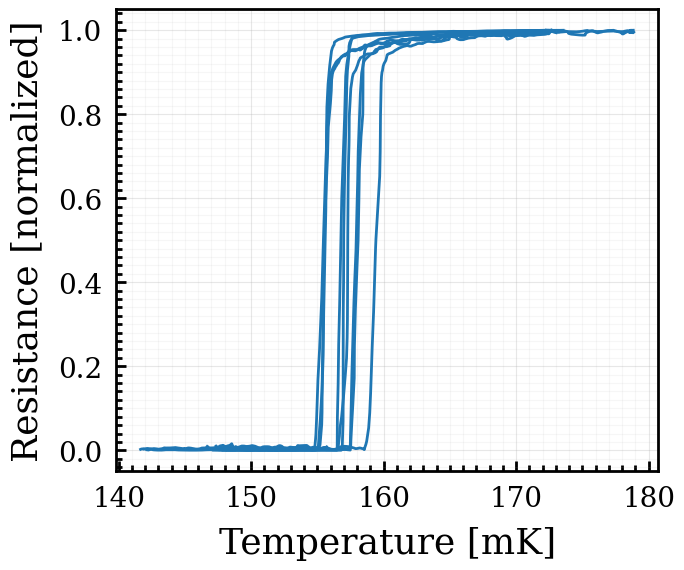}
\includegraphics[width=0.44\textwidth]{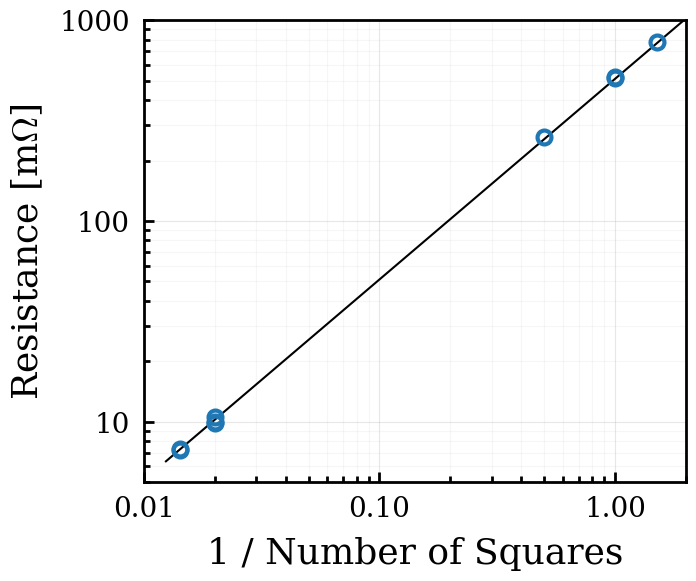}
\caption{\label{fig:TcRn} Four-point resistance measurement. \emph{Left:} Resistance curves for several different interdigitated TES designs across the wafer, normalized to unity. The spread in \Tc\ is $\pm2.5$~mK. \emph{Right:} The measured normal-state resistances is inversely proportional to the number of squares in the interdigitated design, as expected.}
\end{figure}

Four-point resistance measurement of the Hf TESs using the LakeShore resistance bridges are shown in Fig.~\ref{fig:TcRn}. 
We find that the \Tc\ for the patterned bolometer is unchanged from the bare Hf film, confirming that the heated sputter deposition of the Hf is robust against further fab processes. The median measured \Tc\ is 157~mK. The spread in \Tc\ is below $\pm2.5$~mK across the wafer. This spread is similar to that between wafers for a given deposition temperature highlighting the consistency and reliability of Hf as a TES material. 

\begin{figure*}
\includegraphics[width=0.95\textwidth,trim={0cm 6.5cm 0cm 0cm},clip]{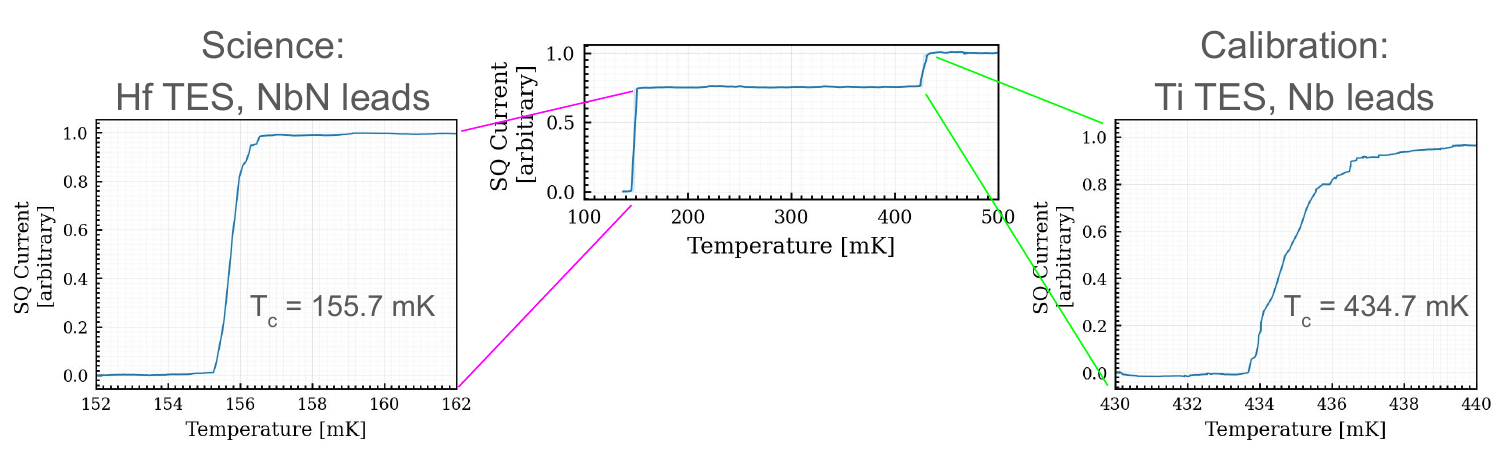}
\caption{\label{fig:Tc_dual} TES resistance as a function of temperature for a dual-\Tc\ design. The transitions at 155.7~mK and 434.7~mK are associated with the Hf-based and Ti-based TES, respectively. Each \Tc\ curve is normalized to unity.}
\end{figure*}

Fig.~\ref{fig:Tc_dual} shows the SQUID current response as a function of bath temperature of a released, dual-\Tc\ TES with both a Hf-based science TES and Ti-based calibration TES located on the same bolometer island (as shown in Fig. \ref{fig:olympus}), transitioning at 155.7~mK and 434.7~mK, respectively. This test structure shows a successful demonstration of the dual-\Tc\ design.


\subsection{Normal-state resistance}


From the four-point resistance measurement the normal-state resistance can also be determined. The plots in Fig.~\ref{fig:TcRn} contain measurements of bolometers with varying $\mathcal{O}(10~\mathrm{m}\Omega)$ and $\mathcal{O}(1~\Omega)$ designs. The ratios between measured normal resistances match the expected ratios from the anticipated $1/N$ scaling, see Table \ref{tab:Tc}. Variability in the normal resistance for a given design is within 8\% of the anticipated normal resistance.

We have successfully used an interdigiated design to reduce the normal resistance from $\mathcal{O}(1~\Omega)$ down to $\mathcal{O}(10~\mathrm{m}\Omega)$ without changes to the fabrication process. This ensures that the Hf TES design and fabrication process is compatible with both TDM, FDM, and $\mu$-mux readout schemes. 
We note that $10~\mathrm{m}\Omega$ and $1~\Omega$ are the bounds we explored here -- any normal resistance between these bounds can be designed. Furthermore, there is no indication that normal resistances outside of these bounds cannot be achieved simply by modifying the interdigitated design, while considering a reasonable TES footprint and the resolution limitation of photo-lithography.

\begin{table}
\caption{\label{tab:Tc}Hf TES bolometers parameters from four-point resistance measurements.}
\begin{ruledtabular}
\begin{tabular}{cccc}
Designed $N$  \footnote{Number of parallel squares in the interdigiated TES design used to reduce $R_{\square}$ by a factor of $N$ to achieve the desired \Rn.} & 
\Rn\ [m$\Omega$] \footnote{$\mathcal{O}(1~\Omega)$ and $\mathcal{O}(10~\mathrm{m}\Omega)$ are separated by a horizontal line.} & 
\Tc\ [mK] & 
$\alpha$\\
\hline
2/3 & 779 & 156 & 315\\
1 & 513 & 157 & 554\\
1 & 523 & 157 & 343\\
2 & 261 & 158 & 291\\ 
2 & 241 & 151 & 299\\\hline
30 & 16.2 & 152 & 339\\
50 & 10.6 & 159 & 253\\
50 & 9.9 & 158 & 336\\
50 & 10.0 & 155 & 308\\
70 & 7.2 & 157 & 336\\
70 & 7.3 & 155 & 282\\
\end{tabular}
\end{ruledtabular}
\end{table}

\subsection{Transition steepness}


During normal bolometer operations, the bolometer is voltage biased such that it is held at a mid-way point in the transition from the superconducting state to the normal state. The term $\alpha=\frac{T}{R}\frac{\mathrm{d} R}{\mathrm{d} T}$ is used to defined the steepness of the transition.\cite{Irwin2005} A steep transition is desired for linear bolometer operations. We find the Hf transition intrinsically occurs over a narrow temperature range, as is seen in Fig.~\ref{fig:TcRn}. We measure $\alpha>200$ consistently, see Table \ref{tab:Tc}.

The steepness of the transition also influences the loop gain, which at DC is defined as $\mathcal{L}=\sfrac{P_{elec}\alpha}{GT_c}$, where $P_{elec}$ if the electrical power applied to the TES bolometer to maintain its voltage bias, and $G$ is the thermal conductance of the bolometer.\cite{Irwin2005} A loop gain greater than 10 is typically desired for CMB experiments. We report on measured loop gain values in Sect. \ref{sec:results:tau}.

\subsection{I-V curves}

\begin{figure*}
\includegraphics[width=0.325\textwidth]{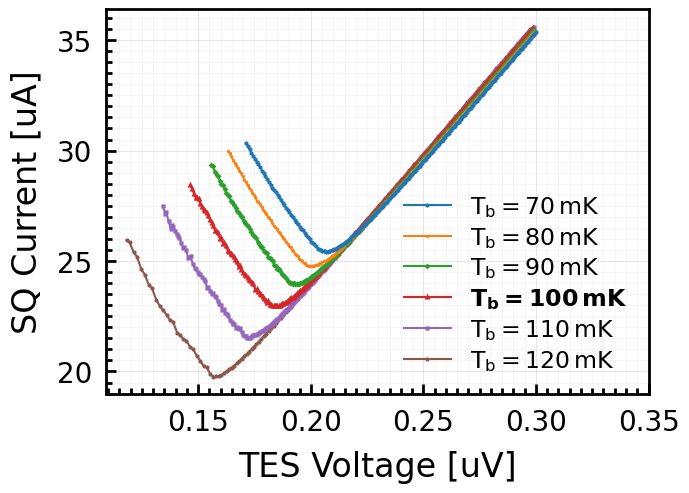}
\includegraphics[width=0.325\textwidth]{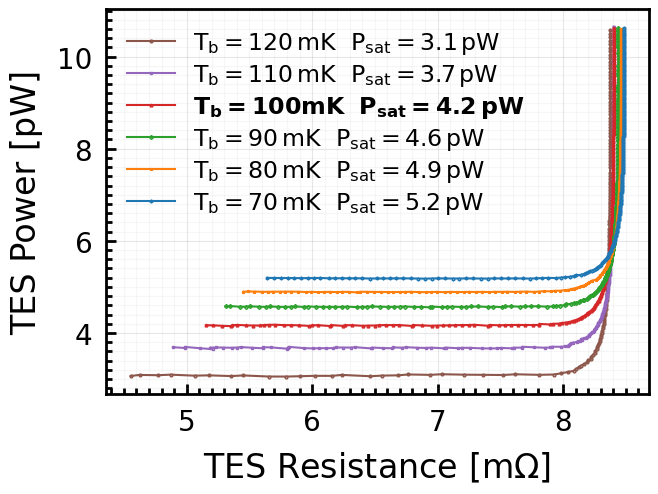}
\includegraphics[width=0.325\textwidth]{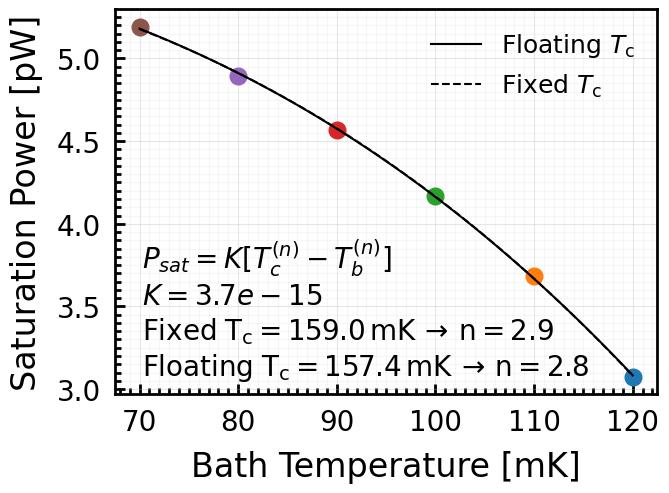}
\caption{\label{fig:IV} \emph{Left:} SQUID current response as a function of TES voltage for varying bath temperatures. The measured TES normal resistance and parasitic resistance are 8.4~m$\Omega$ and 0.77~m$\Omega$, respectively. \emph{Middle:} IV measurement converted to TES power and resistance. The saturation power is measured to be 4.2~pW when $T_{\mathrm{b}}=100$~mK. \emph{Right:} Fitting of \Psat\ as a function of \Tb. $n$ is found to be 2.9 when \Tc\ is fixed at 159~mK (as measured) and 2.8 when \Tc\ is allowed to float.}
\end{figure*}

We measure the SQUID current response as a function of bias voltage (which maps to the current across the TES). Fig.~\ref{fig:IV} are representative plots showing the response of one TES at several different bath temperatures with \Tb$<$\Tc. The sharp turnover in the IV curve is indicative of a steep superconducting transition. The upward trend in the IV curve below the turnover shows that we are able to probe deep into the transition before the TES goes superconducting. 
From the IV curve the saturation power of the TES can be determined. 
We measure \Psat\ to be 4.2~pW when $T_{\mathrm{b}}=100$~mK.
We fit for the relation betwen \Psat\ and \Tb:
$P_{\mathrm{sat}}=K[T_{\mathrm{c}}^{n}-T_{\mathrm{b}}^{n}]$. We consider two scenarios: one in which both \Tc\ and $n$ are free parameters, and one in which we fix \Tc\ to the measured \Tc\ for this TES. The conductivity index $n$ indicates the dominant coupling mechanism. An $n$ of 1 indicates electron coupling while an $n$ of 3 indicates phonon coupling. We fit a conductivity index of 2.9 and 2.8 when \Tc\ is fixed to the measured 159~mK and when \Tc\ is allowed to float, respectively.

\subsection{Time constant and loop gain}\label{sec:results:tau}
The time constant ($\tau$) is a measure of the speed with which a TES recovers after a stimulus. It is given by: $\tau=\tau_0/(\mathcal{L}+1)$, where $\tau_0=C/G$. $C$ is the heat capacity of the TES. We use a large Pd volume to increase the heat capacity. $G$ is the thermal conductivity 
and is inversely proportional to the bolometer leg length. 
When holding the bath temperature constant, the voltage bias applied to the TES sets the temperature and resistance of the TES, which in turn affects the decay time. 
When voltage biasing the bolometer, the loop gain is defined as $\mathcal{L}=\sfrac{\partial P_{\mathrm{bias}}}{\partial P_{\mathrm{total}}}$, from which the previous equation of loop gain stems.\cite {Lee1998} Therefore, if the voltage bias coincides with the IV turnover, slope of the IV curve is zero and the loop gain is equal to unity. Here the intrinsic time constant is then $\tau_0=2\times\tau_{\mathrm{TES}}$.

\begin{figure*}
\includegraphics[width=0.32\textwidth]{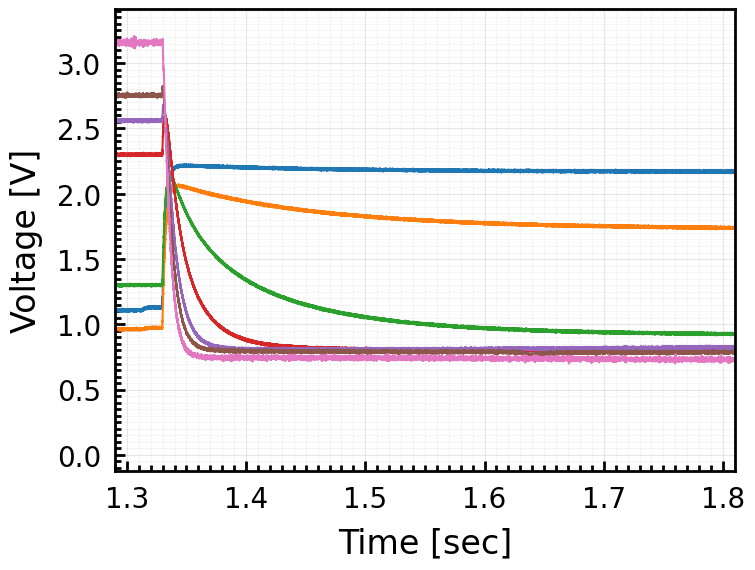}
\includegraphics[width=0.32\textwidth]{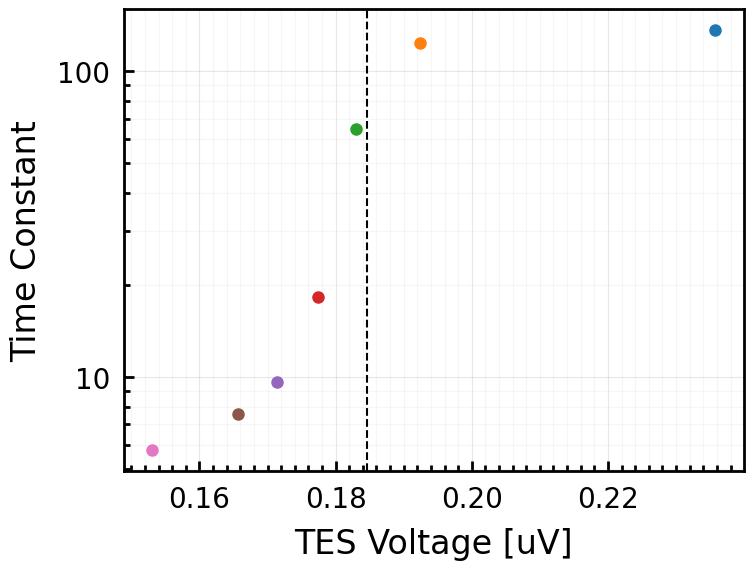}
\includegraphics[width=0.32\textwidth]{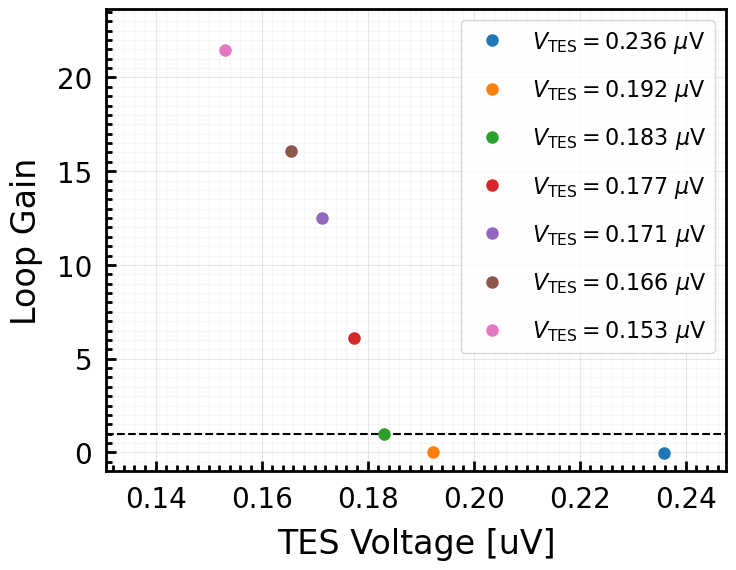}
\caption{\label{fig:timeconst} A 0.5~Hz square signal stimulus is applied to the TES and the SQUID response is recorded for several bias voltages (colored curves, \emph{left}). The time constant is fit for each bias voltage (\emph{middle}) from which a loop gain can be calculated (\emph{right}). The turnover voltage is identified by a vertical dashed line (\emph{middle}) and a loop gain of 1 is included (horizontal dashed line, \emph{right}) to guide the eye.}
\end{figure*}

We apply a square wave stimulus with varying voltage bias (see Fig. \ref{fig:timeconst}) and fit for the decay time in the response. 
The \emph{left} panel in Fig.~\ref{fig:timeconst} shows the falling edge of the TES response to a square wave stimulus. The associated time constants and loop gain are shown in the \emph{middle} and \emph{right} panels, respectively. We measure the loop gain to exceed 10 as we go deeper into the transition. Loop gains greater than 10 are typically desired for CMB experiments.

\section{Conclusion}\label{sec:conclusion}


We have investigated the superconducting properties of hafnium deposited at elevated temperatures and its viability as a TES bolometer for CMB polarimetry experiments. From XRD measurements we conclude that higher deposition temperatures result in 1) larger crystal sizes, 2) lower-stress films, and 3) lower abundances of the m-plane relative to the c-plane $\alpha$ phase, correlating with a lower critical temperature. The empirical linear dependence we observe of \Tc\ on deposition temperature can be used to accurately and reliably target a desired \Tc. The scatter in \Tc\ between wafers at a given deposition temperature and between the Hf film and a patterned bolometer is low at $\pm2.5$~mK. The heated sputter deposition also ensures the \Tc\ is robust against further exposure to heat as long as the initial deposition temperature is not exceeded, an attractive property when compared to other commonly used TES materials such as AlMn.

Hafnium additionally has an intrinsic steep superconducting transition. We measure $\alpha>200$ and find loop gains to exceed 10 deep in the transition. This sharp transition is maintained when NbN is used for the superconducting electrical leads to the TES and it is embedded in the full-stack CMB detector fabrication process. We employ an interdigitated geometry of the NbN leads to precisely design for normal-state resistances ranging from $\mathcal{O}(10~\mathrm{m}\Omega)$ to $\mathcal{O}(1~\Omega)$. In this way, a simple design change enables the Hf-based TES to be compatible with TDM, FDM, and $\mu$-mux readout schemes. We further demonstrate the successful implementation of a dual-\Tc\ bolometer in which a Hf-based TES with NbN leads is used for the science case and a Ti-based TES with Nb leads is used for calibration purposes. 

We have developed a Hf-based TES bolometer for CMB polarimetry experiments, demonstrating its implementation in a full CMB detector stack. 

\begin{acknowledgments}
We wish to acknowledge Finn Babbe (LBNL) and Miguel Daal (UC Santa Barbara) for conducting XRD measurements of our samples and providing insight and analysis. 

This work was supported by the Office of Science, Office of High Energy Physics, of the U.S. Department of Energy under contract No: DE-AC02-05CH11231/B\&R KA2501032: ``Development of high throughput superconducting microfabrication processes for integrated detector module of next-generation Cosmic Microwave Background polarimetry experiment and future high energy physics experiments.''

Work at the Molecular Foundry was supported by the Office of Science, Office of Basic Energy Sciences, of the U.S. Department of Energy under Contract No. DE-AC02-05CH11231.
\end{acknowledgments}

\section*{Data Availability Statement}

The data that support the findings of this study are available from the corresponding author upon reasonable request.




\section*{References}

\bibliography{HfTESbib}

\end{document}